\documentclass[english]{sbrt}
\usepackage[english]{babel}
\usepackage[utf8]{inputenc}

\usepackage[utf8]{inputenc}
\usepackage[T1]{fontenc}
\usepackage[labelfont=bf,font=small]{caption}
\usepackage{graphicx}
\usepackage{listings}
\usepackage{float}
\usepackage{amsmath,amssymb,exscale}
\usepackage{blindtext, graphicx}
\usepackage{verbatim}
\usepackage{algorithm}
\usepackage{algpseudocode}
\usepackage{fancyvrb}
\usepackage{bera}
\usepackage{amsmath}
\usepackage{mathtools}
\usepackage{lipsum}
\usepackage{gensymb}
\usepackage[caption=false]{subfig}
\usepackage{balance}
\usepackage{scalerel}

\usepackage[bookmarks=false]{hyperref}

\def\BibTeX{{\rm B\kern-.05em{\sc i\kern-.025em b}\kern-.08em
		T\kern-.1667em\lower.7ex\hbox{E}\kern-.125emX}}

\makeatletter
\def\BState{\State\hskip-\ALG@thistlm}
\makeatother

\usepackage[%
style=ieee,
backend=biber,
]{biblatex}

\usepackage{xcolor}

\usepackage[nolist,printonlyused]{acronym}      % Acronym

\begin{acronym}
	\acro{3GPP}{3rd generation partnership project}
	\acro{AP}{access point}
	\acro{AWGN}{additive white Gaussian noise}
	\acro{B5G}{beyond fifth generation}
	\acro{BS}{base station}
	\acro{CDF}{cumulative distribution function}
	\acro{CLI}{cross-link interference}
	\acro{CPU}{central processing unit}
	\acro{D-MIMO}{distributed MIMO}
	\acro{DL}{downlink}
	\acro{INI}{inter-network interference}
	\acro{LoS}{line-of-sight}
	\acro{MIMO}{multiple-input and multiple-output}
	\acro{MMSE}{minimum mean square error}
	\acro{NLoS}{non-line of sight}
	\acro{PN}{primary network}
	\acro{SE}{spectral efficiency}
	\acro{SINR}{signal-to-interference-plus-noise ratio}
	\acro{SISO}{single input single output}
	\acro{SLI}{same-link interference}
	\acro{SN}{secondary network}
	\acro{TDD}{time division duplex}
	\acro{UE}{user equipment}
	\acro{UL}{uplink}
	\acro{ULA}{uniform linear array}
\end{acronym}

\usepackage{siunitx}
\newcommand{\SecRef}[2][]{Section#1~\ref{#2}}
\newcommand{\FigRef}[2][]{Fig.#1~\ref{#2}}
\newcommand{\TabRef}[2][]{Table#1~\ref{#2}}

\def\BibTeX{{\rm B\kern-.05em{\sc i\kern-.025em b}\kern-.08em
		T\kern-.1667em\lower.7ex\hbox{E}\kern-.125emX}}

\begin{document}

\title{Study on coexisting 6G D-MIMO and massive MIMO networks operating in TDD}

\author{Maria Clara R. Lobão , Yuri C. B. Silva, Igor M. Guerreiro, Roberto P. Antonioli, Behrooz Makki
\thanks{This work was supported in part by Ericsson Research, Sweden, and Ericsson Innovation Center, Brazil, Technical Coop. Contracts UFC.53 and UFC.55, in part by CNPq, in part by CNPq/INCT-Signals Grant 406517/2022-3, in part by CAPES - Finance Code 001.}%
\thanks{Maria Clara R. Lobão , Yuri C. B. Silva, Igor M. Guerreiro, Roberto P. Antonioli are with the Wireless Telecommunications Research Group (GTEL), Federal University of Cear\'a (UFC), Fortaleza, Brazil. E-mails: \{clara, yuri, igor, antonioli\}@gtel.ufc.br.
	Behrooz Makki is with Ericsson Research, Ericsson AB, Gothenburg, Sweden. (e-mail: behrooz.makki@ericsson.com).
}%
}

\maketitle

% \markboth{XLIV BRAZILIAN SYMPOSIUM ON TELECOMMUNICATIONS AND SIGNAL PROCESSING - SBrT 2026, SEPTEMBER 29TH TO OCTOBER 2ND, 2026, SALVADOR, BA}{}

\begin{abstract}
In this study, the coexistence between a \ac{MIMO} \ac{PN} and a \ac{D-MIMO} \ac{SN} is analyzed when both networks operate in \ac{TDD} and their \ac{UL} and \ac{DL} phases are not synchronized. The impact of the interference is observed when the networks operate with different 3GPP slot patterns, which vary in terms of UL/DL proportion and slot position, resulting in a significant impact on their overall \ac{SE}. Given that, we propose a slot pattern coordination to reduce the perceived interference on the \ac{PN} network by controlling the features of the pattern used by the  \ac{SN}.  The results show that this strategy helps to avoid loss of performance at the \ac{PN}.
\end{abstract}

\section{Introduction}
\acresetall
The increasing demand for higher performance and ubiquituous connection pushes the advance of mobile network technologies \ac{B5G}, which must ensure better data traffic rates, lower latencies and reliable communication. The distributed \acl{MIMO} (\acs{D-MIMO}) networks emerge in this context \cite{Matthaiou2021}, with its user-centric design, which eliminates cell borders for the \ac{UE} \cite{Ngo2017, found,Bjornson2019}, and its cooperative service among \acp{AP}, providing a more uniform coverage and better interference mitigation.

\acused{MIMO}\acused{D-MIMO}

The use of more frequency has become adamant to attend to the growing number of \acp{UE} and the higher rates appeal, however, the availability of adequate bands in the spectrum is not guaranteed, as it is expensive, mostly occupied and a limited resource \cite{Yang2019}. Therefore, in order to meet these demands, the implementation of new networks may require their coexistence with other networks that already make use of the required frequency bands.

Recent works have tackled the coexistence topic considering the most varied scenarios. In \cite{Shaik2024a}, the authors consider the coexistence of a \ac{D-MIMO} \ac{SN} and a \ac{MIMO} \ac{PN}. In \cite{Elfi2023}, the coexistence between a cellular \ac{MIMO} network and a \ac{MIMO} radar system is explored.

In the aforementioned works, as well as most coexistence studies, it is assumed that the link direction of the \ac{PN} and \ac{SN} are completely synchronized, i.e., the \ac{UL} and \ac{DL} of both systems occur at the same time. This means that the only type of interference considered in these analyses is the \ac{SLI}. In practical scenarios, however, some level of asynchonism is expected for networks operating in \ac{TDD}, which leads to the occurrence of \ac{CLI}. % 

In this work, we investigate the effects of both \ac{SLI} and \ac{CLI} in a \ac{TDD} co-channel coexistence scenario where the \ac{PN} is a cellular \ac{MIMO} and the \ac{SN} is a \ac{D-MIMO}. A dynamic framework is investigated, where both networks operate with different slot patterns based on the \ac{3GPP} configuration \cite{3gpp.38.213}, and the \ac{SE} performance of the \ac{PN} is measured for every combination of patterns with the \ac{SN}. Moreover, a slot coordination is proposed, where the \ac{PN} indicates to the \ac{SN} certain proportions and positions of \ac{UL} or \ac{DL} slots that can be employed based on the current slot pattern that is being used by the \ac{PN}. The results show that this policy can improve the performance of the worst or the best \acp{UE} depending on the suggested configuration.

\section{System Model}\label{System_Model}

We consider the \ac{UL} and the \ac{DL} of a \ac{D-MIMO} \ac{SN} and a massive \ac{MIMO} \ac{PN} network coexisting in the same frequency resource with \ac{INI}. The \ac{SN} consists of $L$ \acp{AP}, each equipped with $N$ antennas, which simultaneously serve $K$ single-antenna \acp{UE}. The \acp{AP} are connected via an ideal backhaul link to a \ac{CPU}. The \ac{PN} is a single cell with one \ac{BS} equipped with $P$ antennas serving $M$ single-antenna \acp{UE}.

\begin{figure}[h!]
	\centering
	\includegraphics[width=\linewidth]{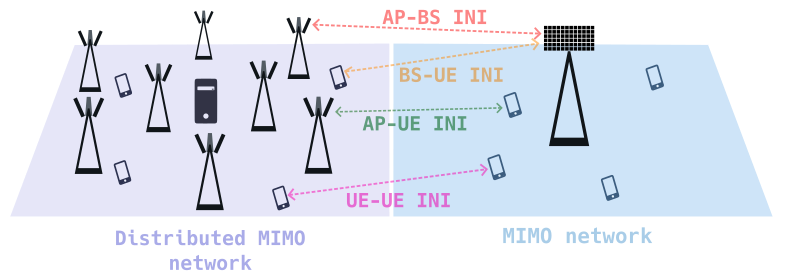}
	\caption{Illustration of the scenario with all the types of \ac{INI} that can occur.}
	\label{fig:sysfig}
\end{figure}

The networks operate in \ac{TDD} and the length of the coherence block is the same value of $\tau_c$ for both, with $\tau_{u}$ and $\tau_{d}$ being the length of the \ac{UL} and \ac{DL} phases, respectively. We consider that all symbols in the slot have equal length in both networks, therefore, the switch from \ac{UL} to \ac{DL} occurs at the same time in the \ac{PN} and in the \ac{SN}. On the other hand, the progression of the slots is distinct for each network, so in each slot the systems can have their phases synchronized (i.e., \ac{UL}/\ac{UL} or \ac{DL}/\ac{DL}), causing \ac{SLI}, or unsynchronized (i.e, \ac{UL}/\ac{DL}), causing \ac{CLI}. Given all the \ac{SLI} and \ac{CLI} cases, we define next the channels that exist between all combinations of transmitting and receiving agents in this coexistence scenario, which are shown in \FigRef{fig:sysfig}. 

The channel between a \ac{UE} $i$ and a receiving array Rx, that can be either an \ac{AP} or the \ac{BS}, follows a Rician model described as:
\begin{equation}
	\mathbf{h}_{i,\text{Rx}} = \left(\sqrt{\cfrac{\Bar{K}_{i,\text{Rx}}}{\Bar{K}_{i,\text{Rx}}+1}}\mathbf{a}_{i,\text{Rx}} + \sqrt{\cfrac{1}{\Bar{K}_{i,\text{Rx}}+1}}\mathbf{h}^{(w)}_{i,\text{Rx}}\right)\sqrt{\beta_{i,\text{Rx}}}  \, ,
	\label{eq:ricianFadingChannel}
\end{equation}
where $\Bar{K}_{i,\text{Rx}}$ is the Rician K-factor between the \ac{UE} and the receiver, $\mathbf{h}^{(w)}_{i,\text{Rx}}\sim \mathcal{CN}\left(0,\mathbf{R}_{i,\text{Rx}}\right)$ is the fast-fading \ac{NLoS} component, $\mathbf{R}_{i,\text{Rx}}$ is the spatial correlation matrix, and $\mathbf{a}_{i,\text{Rx}}$ is the steering vector of the receive \ac{ULA}. The term $\beta_{i,\text{Rx}}$ is the large scale fading coefficient, which is given by:
\begin{equation}
	\beta_{i,\text{Rx}} = 10^{\frac{\textrm{PL}_{i,\text{Rx}}+\textrm{SH}_{i,\text{Rx}}}{10}} \, ,
	\label{eq:LSFcoeffic}
\end{equation}
in which $\textrm{PL}_{i,\text{Rx}}$ is the path-loss and $\textrm{SH}_{i,\text{Rx}}$ is the shadowing, all in logarithmic scale. The propagation parameters such as path-loss, probability of \ac{LoS}, shadowing standard deviaton $\sigma_{sh}$ and Rician K-factor $\Bar{K}_{i,\text{Rx}}$ were computed following the \ac{3GPP} report 38.858 \cite{3gpp.38.858}, which is an adaptation of the report 38.901 \cite{3gpp.38.901} for \ac{TDD} scenarios. The report \cite{3gpp.38.858} provides the propagation attributes for links between \acp{UE} and between \acp{BS} (or \acp{AP}).

Since \eqref{eq:ricianFadingChannel} represents the channel between a \ac{UE} and a generic receiver, the dimensions of the vectors and matrices will depend on which receiver is receiving the signal. If it is an \ac{SN} \ac{AP}, then $\mathbf{h}_{i,\text{Rx}} \in \mathbb{C}^{N}$, $\mathbf{h}^{(w)}_{i,\text{Rx}} \in \mathbb{C}^{N}$, $\mathbf{a}_{i,\text{Rx}} \in \mathbb{C}^{N}$ and $\mathbf{R}_{i,\text{Rx}} \in \mathbb{C}^{N \times N}$. If it is a \ac{PN} \ac{BS}, then $\mathbf{h}_{i,\text{Rx}} \in \mathbb{C}^{P}$, $\mathbf{h}^{(w)}_{i,\text{Rx}} \in \mathbb{C}^{P}$, $\mathbf{a}_{i,\text{Rx}} \in \mathbb{C}^{P}$ and $\mathbf{R}_{i,\text{Rx}} \in \mathbb{C}^{P \times P}$. 

Considering the cross-channels of the systems, that is, the links \ac{UE}-\ac{UE} and \ac{AP}-\ac{BS}, the expression in \eqref{eq:ricianFadingChannel} suffers some changes. When the link is between two \acp{UE} it becomes a \ac{SISO} channel expressed as:
\begin{equation}
	h_{i,j} = \left(\sqrt{\cfrac{\Bar{K}_{i,j}}{\Bar{K}_{i,j}+1}} + \sqrt{\cfrac{1}{\Bar{K}_{i,j}+1}}h^{(w)}_{i,j}\right)\sqrt{\beta_{i,j}}  \, ,
	\label{eq:ricianFadingChannelUE2UE}
\end{equation}
with the \ac{NLoS} component $h^{(w)}_{i,j} \sim \mathcal{CN}(0,1)$. When the link is from the \ac{BS} to an \ac{AP}, the channel is \ac{MIMO} and \eqref{eq:ricianFadingChannel} becomes the matrix $\mathbf{H}_{\text{BS},l} \in \mathbb{C}^{N \times P}$, with $l = 1, \dots, L$ being the \ac{AP} index, described as:
\begin{equation}
	\mathbf{H}_{\text{BS},l} = \left(\sqrt{\cfrac{\Bar{K}_{\text{BS},l}}{\Bar{K}_{\text{BS},l}+1}}\mathbf{A}_{\text{BS},l} + \sqrt{\cfrac{1}{\Bar{K}_{\text{BS},l}+1}}\mathbf{H}^{(w)}_{\text{BS},l}\right)\sqrt{\beta_{\text{BS},l}}  \, ,
	\label{eq:ricianFadingChannelAP2BS}
\end{equation}
where $\mathbf{A}_{\text{BS},l} \in \mathbb{C}^{N \times P}$ is the steering matrix, $\mathbf{H}^{(w)}_{\text{BS},l} \in \mathbb{C}^{N \times P} \sim \mathcal{CN}(0,\mathbf{R}_{\text{BS},l})$ is the \ac{NLoS} fast-fading component, and $\mathbf{R}_{\text{BS},l} \in \mathbb{C}^{N \times N}$ is the correlation matrix. When the channel is from the \ac{AP} $l$ to the \ac{BS}, we have $\mathbf{H}_{l,\text{BS}} \in \mathbb{C}^{P \times N}$, computed analogously.

\section{Signal Model}\label{Signal_Model}

As described in \SecRef{System_Model}, the \ac{UL} and \ac{DL} slot dynamics between the coexisting networks generates different types of interference. In this section, we present the signal model for the \ac{UL} and \ac{DL} of each network when they are under \ac{SLI} and \ac{CLI}. 

The \ac{D-MIMO} \ac{SN} is considered to be operating in the fully centralized mode \cite{found}, then the vectors are shown in a compacted form, such as:
\begin{equation}
	\mathbf{h}_k = \begin{bmatrix}
		\mathbf{h}_{k,1}^T &
		\mathbf{h}_{k,2}^T &
		\cdots &
		\mathbf{h}_{k,L}^T
	\end{bmatrix}^T \, , 
\end{equation}
where $\mathbf{h}_k \in \mathbb{C}^{LN}$ is the collective channel vector. The same is done for the collective combining vector $\mathbf{v}_k \in \mathbb{C}^{LN}$, and the collective precoding vector $\mathbf{w}_k \in \mathbb{C}^{LN}$.
\subsection{Primary Network}

In order to distinguish the vectors from the \ac{PN} to the ones from the \ac{SN}, all vectors related to elements from the \ac{PN} are marked with $(\cdot)^{\star}$.

When the system is in the \ac{UL}, the signal received by the \ac{BS} is the combination of all the signals sent by the $M$ \acp{UE} plus the interference from the \ac{SN}. Let $u_s$ be the binary variable that determines the direction in which the network operates in the symbol $s = 1,2,\dots,S$ of the slot, if $s$ is in \ac{DL}, $u_s = 0$, otherwise $u_s=1$. Considering that the \ac{PN} is in the \ac{UL} in $s$, the combined received signal vector $	\left(\mathbf{y}_s^{\star}\right)^{\text{UL}} \in \mathbb{C}^{P}$ is defined as:
\begin{equation}
	(\mathbf{v}^\star_m)^H(\mathbf{y}_s^{\star})^{\text{UL}} = \sum_{i=1}^{M} (\mathbf{v}^\star_m)^H\mathbf{h^\star}_i \zeta^\star_i +   I^{\star}_{s,\text{UL}}+ (\mathbf{v}^\star_m)^H\mathbf{n}^\star \, ,
	\label{eq:ULPrimarySignal}
\end{equation}
where $\mathbf{h^\star}_i \in \mathbb{C}^{P}$ is the channel between the \ac{PN} \ac{UE} $i$ and the \ac{BS}, following the model in \eqref{eq:ricianFadingChannel}. The term $\mathbf{n}^\star \in \mathbb{C}^{P}$ is the \ac{AWGN} at the receiver and $\zeta^\star$ is the \ac{UL} complex signal sent by the \ac{PN} \ac{UE}. The vector $\mathbf{v}^\star_m \in \mathbb{C}^{P}$ is the combiner used to extract the signal $\zeta^\star_m$, where $m=1,\dots,M$ is the desired \ac{UE} index. The term $I^{\star}_{s,\text{UL}}$ represents the perceived interference from the \ac{SN} at the \ac{UL} signal of the primary, described by the equation:
\begin{equation}
	I^{\star}_{s,\text{UL}} = u_s\sum_{j=1}^{K} (\mathbf{v}^\star_m)^H\mathbf{h}_{j}\zeta_j + (1-u_s)\sum_{j=1}^{K} (\mathbf{v}^\star_m)^H\mathbf{H}_{\text{AP}}\mathbf{w}_j\xi_j  \, ,
	\label{eq:interfSecUlPrim}
\end{equation}
where $\mathbf{h}_{j} \in \mathbb{C}^{P}$ is the channel between the \ac{SN} \ac{UE} $j$ and the \ac{BS}, also following \eqref{eq:ricianFadingChannel},  $\mathbf{w}_j \in \mathbb{C}^{LN}$ is the precoding vector from all \acp{AP} to \ac{UE} $j$, $\zeta_j$ is the \ac{UL} signal sent by \ac{UE} $j$, $\xi_j$ is the \ac{DL} signal sent from the \acp{AP} to \ac{UE} $j$ and $\mathbf{H}_{\text{AP}} \in \mathbb{C}^{P \times LN}$ is the channel matrix from all \acp{AP} to the \ac{BS} defined as 
$
\mathbf{H}_{\text{AP}} = \begin{bmatrix}
	\mathbf{H}_{1,\text{BS}} &
	\mathbf{H}_{2,\text{BS}} &
	\cdots &
	\mathbf{H}_{L,\text{BS}}
\end{bmatrix}
$.

In this case, the \ac{SINR} for \ac{UE} $m$ is computed according to\cite{found}:
\begin{align}
	&\text{SINR}^{\text{UL}}_{s,m} = p^\star_m|\mathbb{E}\{(\mathbf{v}^\star_m)^H \mathbf{h}^\star_m\}|^2\Bigg(\sum_{i=1}^{M}p^\star_i\mathbb{E}\{|(\mathbf{v}^\star_m)^H\mathbf{h}^\star_i|^2\} - \nonumber \\ 
	&p^\star_m|\mathbb{E}\{(\mathbf{v}^\star_m)^H \mathbf{h}^\star_m\}|^2 + \mathbb{E}\{|I^{\star}_{s,\text{UL}}|^2\} + \sigma^2_{\text{UL}}\mathbb{E}\{||(\mathbf{v}^\star_m)^H||^2\}\Bigg)^{-1}\, .
	\label{sinrULPN}
\end{align}
The expectations are taken on the channel realizations and $p_i = \mathbb{E}\{|\zeta_i|^2\}$.

If the \ac{PN} is in the \ac{DL} in $s$, the signal received by the \ac{UE} $m$ is a scalar $(y^{\star}_m)^\text{DL}$, which is a result of the combination of signals sent from the \ac{BS} to all \acp{UE} precoded by the vector $\mathbf{w}^\star_m \in \mathbb{C}^P$, plus the interference arriving from the \ac{D-MIMO} system. In this case, the received signal is:
\begin{align}
	&(y^{\star}_{s,m})^{\text{DL}} = \sum_{i=1}^{M} (\mathbf{h}_{m}^\star)^H\mathbf{w}^\star_{i}\xi^\star_i  + I^{\star}_{s,\text{DL}} + n_m \, ,\\
	&I^{\star}_{s,\text{DL}} = u_s\sum_{j=1}^K h_{m,j}\zeta_j  + (1-u_s)\sum_{j=1}^K \mathbf{g}_{m}^H\mathbf{w}_j\xi_j \, ,
\end{align}
where $n_m$ is the \ac{AWGN} at the receiver and $\xi^\star_i$ is the \ac{DL} symbol received by the \ac{PN} \ac{UE} from the \ac{BS}.
$h_{m,j}$ is the channel from the \ac{SN} \ac{UE} $j$ to the \ac{PN} \ac{UE} $m$ and $\mathbf{g}_{m} \in \mathbb{C}^{LN}$ is the channel between the \ac{UE} $m$ and all the \ac{SN} \acp{AP}. This expression leads to the \ac{SINR}:
\begin{align}
	\text{SINR}^{\text{DL}}_{s,m} &= |\mathbb{E}\{(\mathbf{h}^\star_m)^H \mathbf{w}^\star_m\}|^2\Bigg(\sum_{i=1}^M \mathbb{E}\{|(\mathbf{h}^\star_m)^H\mathbf{w}^\star_i|^2\} - \nonumber \\
	& |\mathbb{E}\{(\mathbf{h}^\star_m)^H \mathbf{w}^\star_m\}|^2 + \mathbb{E}\{|I^{\star}_{s,\text{DL}}|^2\} + \sigma_\text{DL}^2\Bigg)^{-1}\, .
	\label{sinrDLPN}
\end{align}

\subsection{Secondary Network}

The signal model of the \ac{D-MIMO} is also based on \cite{found} and \cite{mMIMO2017}. If the \ac{SN} is operating in \ac{UL} in $s$, the signal $\mathbf{y} \in \mathbb{C}^{LN}$ received by the \acp{AP} and combined with the vector $\mathbf{v}_k$ for the desired signal of the \ac{UE} $k$ is:
\begin{align}
	&\mathbf{v}^H_k\mathbf{y}_s^{\text{UL}} = \sum_{i=1}^{K} \mathbf{v}^H_k\mathbf{h}_i \zeta_i + I_{s,\text{UL}} + \mathbf{v}^H_k\mathbf{n} \, , \label{eq:receivedSig} \\
	&I_{s,\text{UL}} = u_s\sum_{j=1}^{M} \mathbf{v}^H_k\mathbf{h}^\star_{j}\zeta^\star_j  + (1-u_s)\sum_{j=1}^{M} \mathbf{v}^H_k\mathbf{H}^\star_{\text{BS}}\mathbf{w}^\star_j\xi^\star_j \, ,
\end{align}
where $\mathbf{n} \in \mathbb{C}^{LN}$ is the \ac{AWGN} at the receiving \acp{AP}. The matrix $\mathbf{H}^\star_{\text{BS}} \in \mathbb{C}^{LN \times P}$ is the channel matrix from the \ac{PN} \ac{BS} to all \acp{AP}, which is modeled as 
$
\mathbf{H}^\star_{\text{BS}} = \begin{bmatrix}
	\mathbf{H}^T_{\text{BS},1} &
	\mathbf{H}^T_{\text{BS},2} &
	\cdots &
	\mathbf{H}^T_{\text{BS},L}
\end{bmatrix}^T \, .
$
This leads to:
\begin{align}
	&\text{SINR}^{\text{UL}}_{s,k} = p_k|\mathbb{E}\{\mathbf{v}^H_k
	\mathbf{h}_k\}|^2 \Bigg(\sum_{i=1}^{K}p_i\mathbb{E}\{|\mathbf{v}^H_k 
	\mathbf{h}_i|^2\} -  \nonumber \\
	&p_k|\mathbb{E}\{\mathbf{v}^H_k 
	\mathbf{h}_k\}|^2 +\mathbb{E}\{|I_{s,\text{UL}}|^2\} + \sigma^2_{\text{UL}}\mathbb{E}\{||\mathbf{v}^H_k
	||^2\}\Bigg)^{-1}\, .
	\label{sinrULSN}
\end{align}
For the \ac{DL} of the \ac{SN}, the signal received by \ac{UE} $k$ is:
\begin{align}
	&y_{s,k}^{\text{DL}} = \sum_{i=1}^{K} \mathbf{h}_{k}^H
	\mathbf{w}_{i}\xi_i + I_{s,\text{DL}} + n_k\, ,\\
	&I_{s,\text{DL}} = u_s\sum_{j=1}^M h_{k,j}^\star \zeta^\star_j + (1-u_s)\sum_{j=1}^M (\mathbf{g}^\star_{k})^H\mathbf{w}^\star_j\xi^\star_j\, ,
\end{align}
in which $h_{k,j}^\star$ is the channel from the \ac{PN} \ac{UE} $j$ to \ac{SN} \ac{UE} $k$ and $\mathbf{g}^\star_{k} \in \mathbb{C}^{P}$ is the channel between the \ac{UE} $k$ and the \ac{BS}. For this case, the \ac{SINR} is expressed as:
\begin{align}
	\text{SINR}^{\text{DL}}_{s,k} &= |\mathbb{E}\{\mathbf{h}^H_k\mathbf{w}_k\}|^2\Big(\sum_{i=1}^K \mathbb{E}\{|\mathbf{h}^H_k\mathbf{w}_i|^2\} - \nonumber \\ &|\mathbb{E}\{\mathbf{h}^H_k\mathbf{w}_k\}|^2 + \mathbb{E}\{|I_{s,\text{DL}}|^2\} + \sigma_\text{DL}^2\Big)^{-1}\, .
	\label{sinrDLSN}
\end{align}

In this scenario, the channel estimation is based on the transmission of orthogonal pilot sequences of same length $\tau_p$. The estimates are computed via a \ac{MMSE} estimator as in \cite{found}. %
We assume that the channel estimation phase is simultanous and optimized in a way that the interference from the other system is negligible. 

Moreover, the combiners $\mathbf{v}_i$ and precoders $\mathbf{w}_i$ are computed also following the \ac{MMSE} models in \cite{mMIMO2017} for the \ac{PN} and in \cite{found} for the \ac{SN}.

\section{Slot Pattern Coordination}\label{strategy}

Several combinations of slot patterns can take place between the \ac{PN} and \ac{SN}, given the configurations available in \cite{3gpp.38.213}. These combinations can result in a significant impact in terms of interference in both networks due to the quantity of \ac{DL}/\ac{UL} slots present in each pattern, as well as how these slots are distributed. Depending on how strong the \ac{CLI} and \ac{SLI} are in these combinations, the coexisting networks can mutually cause great impact on their per-user performance. 

In this work, we propose a simple coordination strategy in order to mitigate this impact, focusing on the \ac{PN}. This strategy relies on the assumption of a certain level of communication between the networks and a limited number of slot patterns available for them. We also assume that the networks have knowledge of their respective performance behavior when submitted to each combination, due to estimates or measurements. 

\begin{figure}[h!]
	\centering
	\includegraphics[width=\linewidth]{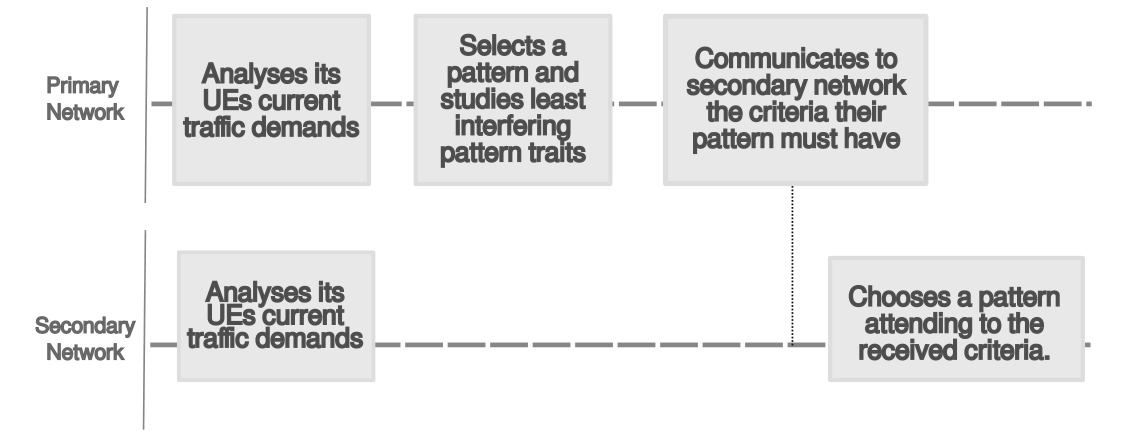}
	\caption{Slot coordination scheme.}
	\label{fig:slotflux}
\end{figure}

The strategy works as follows: the \ac{PN} selects the slot pattern to be employed, based on its \acp{UE}' demands. Fixing the chosen pattern, it then analyzes the aspects of the \ac{SN} pattern, i.e., \ac{DL}/\ac{UL} slot positions and proportions, that would cause the least impact on subsequent \ac{PN} transmissions. Given this analysis, it communicates instructions to the \ac{SN}, for example, if the \ac{SN} pattern must start with \ac{DL} slots, or the minimal proportion of \ac{UL} slots the configuration must have. The \ac{SN} then chooses its pattern based on these intructions and on its \acp{UE}' necessities. \FigRef{fig:slotflux} summarizes the scheme.

\section{Simulation Results}\label{Results}

The results are obtained via the \ac{CDF} of the per-user \ac{SE}. The \ac{SE} for the \ac{UL} and \ac{DL} of the systems is computed following the expression, for any $s$:
\begin{equation}
	\text{SE}_i^{\text{UL/DL}} = (\tau_{u/d}/\tau_c)\log_2(1+\text{SINR}^{\text{UL/DL}}_i) \, .
\end{equation}
We considered a horizontal gap of $100$ m between the networks. The statistics were taken over $200$ channel realizations and $500$ Monte Carlo iterations. Other parameters and dimensions are presented in \TabRef{table:parameters}.
\begin{table}[t]
	\centering
	\caption{Simulation parameters.}
	\begin{tabular}{l c l}
		\hline 
		\textbf{Parameter} &&  \textbf{Value} \\
		\hline 
		Carrier frequency &&  $f_c$ = 2 GHz \\
		Bandwidth && $W$ = 20 MHz\\
		Number of \acp{UE} && $M = K = 10$ \\
		Number of \acp{AP} && $L = 16$ \\ 
		Number of antennas && $P=64$, $N=4$\\
		Coverage area for \ac{PN}, \ac{SN} && 0.25 km$^2$,  0.25 km$^2$\\
		\ac{UE}, \ac{AP} and \ac{BS} heights && $h_{\text{UE}}$ = 1.5 m, $h_{\textrm{AP}}$ = $h_{\textrm{BS}}$ = 10 m \\
		\ac{UL} power per \ac{UE}s && $p_k$ = $p_m$ = 100 mW \\
		\ac{DL} powers && $\rho_m$ = 2.5 W,  $\rho_k$ = 200 mW\\
		Antenna spacing && $d$ = $\left(1/2\right)\lambda$ \\
		Number of pilots && $\tau_p$ = 10 \\
		Coherence block length && $\tau_c$ = 200 \\
		\hline
	\end{tabular}
	\label{table:parameters}
\end{table}

\begin{figure}[t]
	\centering
	\vspace{-0.3cm}
	\includegraphics[width=\linewidth]{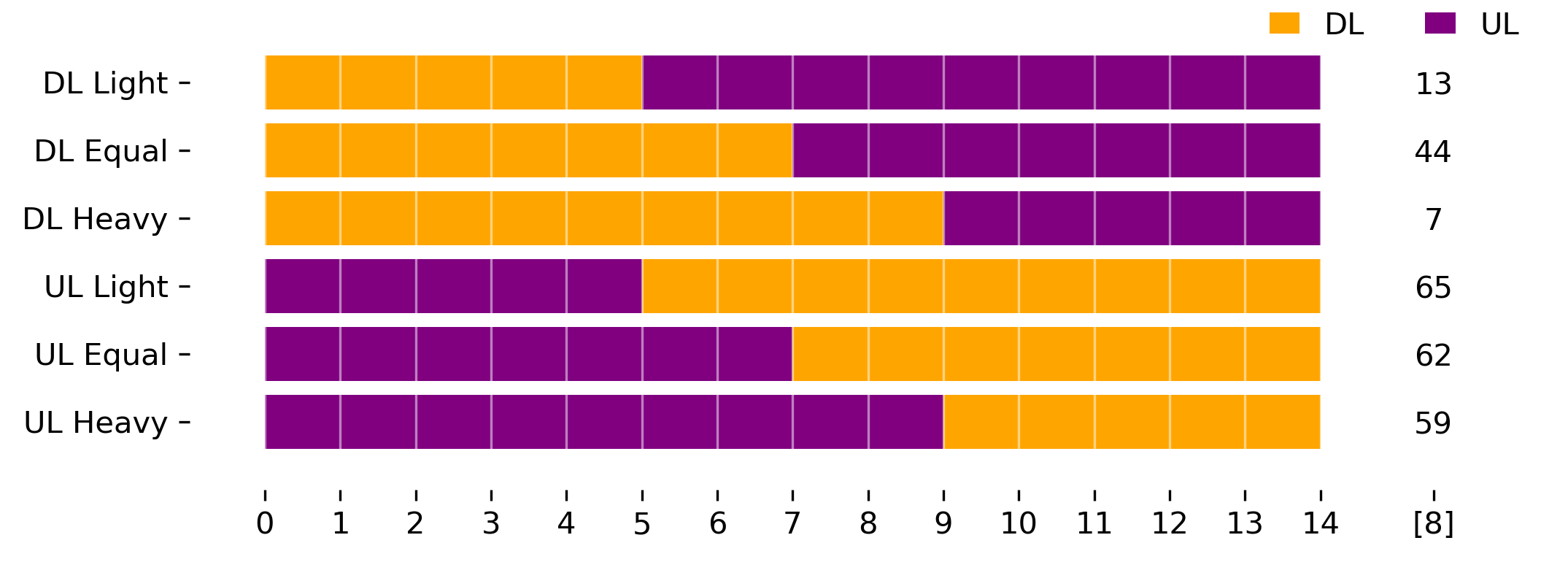}
	\caption{Slot formats and their identification in \cite{3gpp.38.213}.}
	\label{fig:pat}
\end{figure}

For the analysis of the interference dynamics when the networks operate with different slot patterns, we focused on the results of the \ac{PN}. The patterns were selected with the objective to show the influence of the proportion of \ac{DL} and \ac{UL} slots as well as the type of interference that occurs. \FigRef{fig:pat} shows the patterns that were used in this work, which were adapted from the patterns defined in the 3GPP technical specification 38.213 \cite{3gpp.38.213}; the figure shows the original identification number of the patterns in \cite{3gpp.38.213}. 

\begin{figure}[h!]
	\centering
	\includegraphics[width=\linewidth]{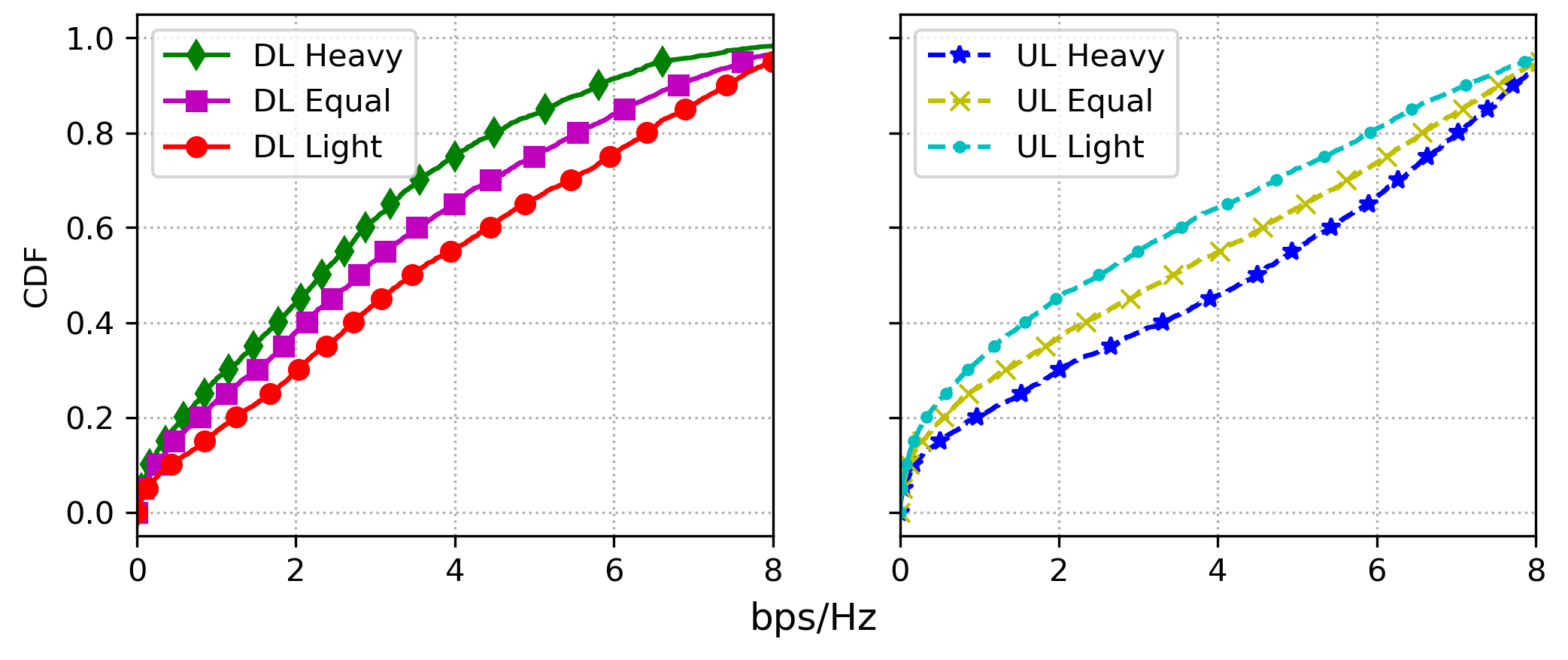}
	\caption{\ac{SE} \ac{CDF} of the \ac{PN} using the DL Heavy pattern varying the pattern used by the \ac{SN}.}
	\label{fig:sepndlhcomp}
	\vspace{-0.5cm}
\end{figure}

\begin{figure}[h!]
	\centering
	\includegraphics[width=.49\linewidth]{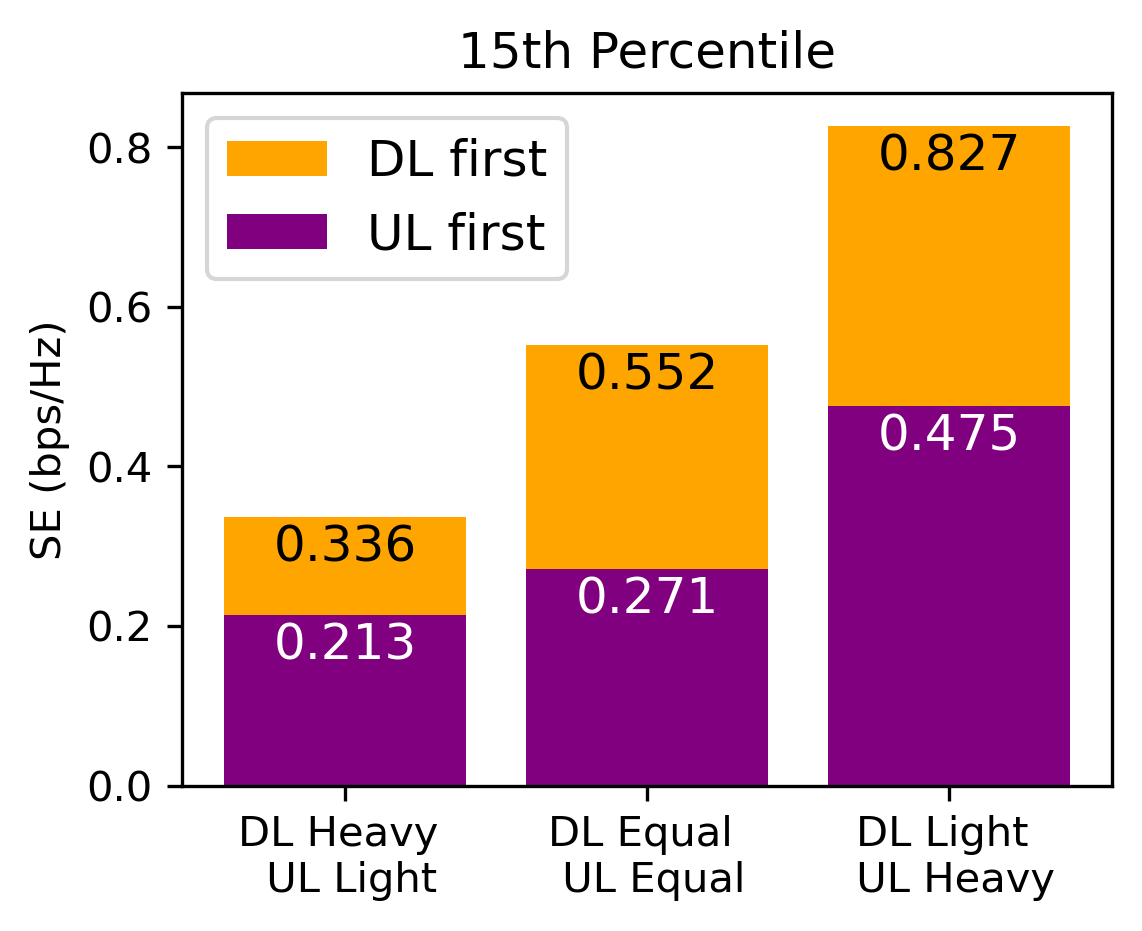}
	\includegraphics[width=.49\linewidth]{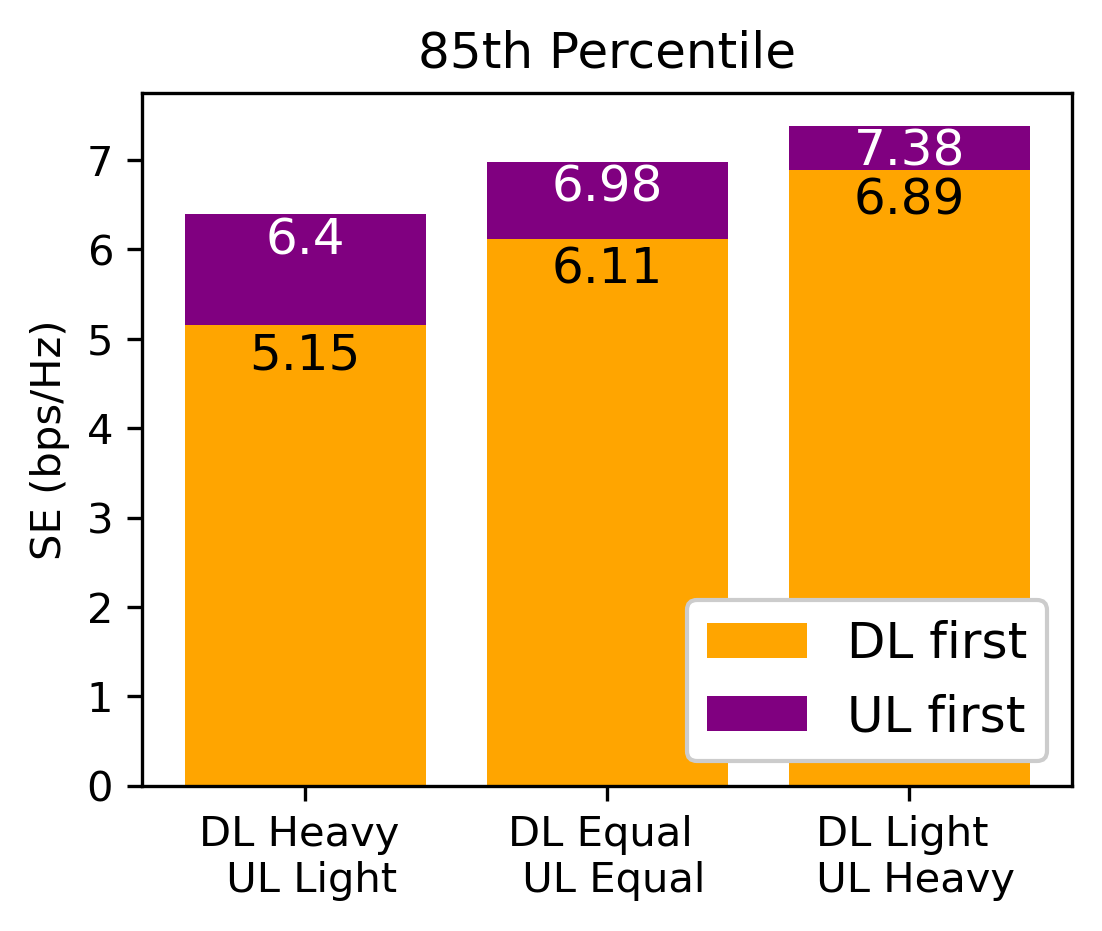}
	\caption{\ac{SE} of the \ac{PN} in \FigRef{fig:sepndlhcomp} in the 15th and 85th percentiles.}
	\label{fig:comp}
	\vspace{-0.3cm}
\end{figure}

\begin{figure}[h!]
	\centering
	\includegraphics[width=\linewidth]{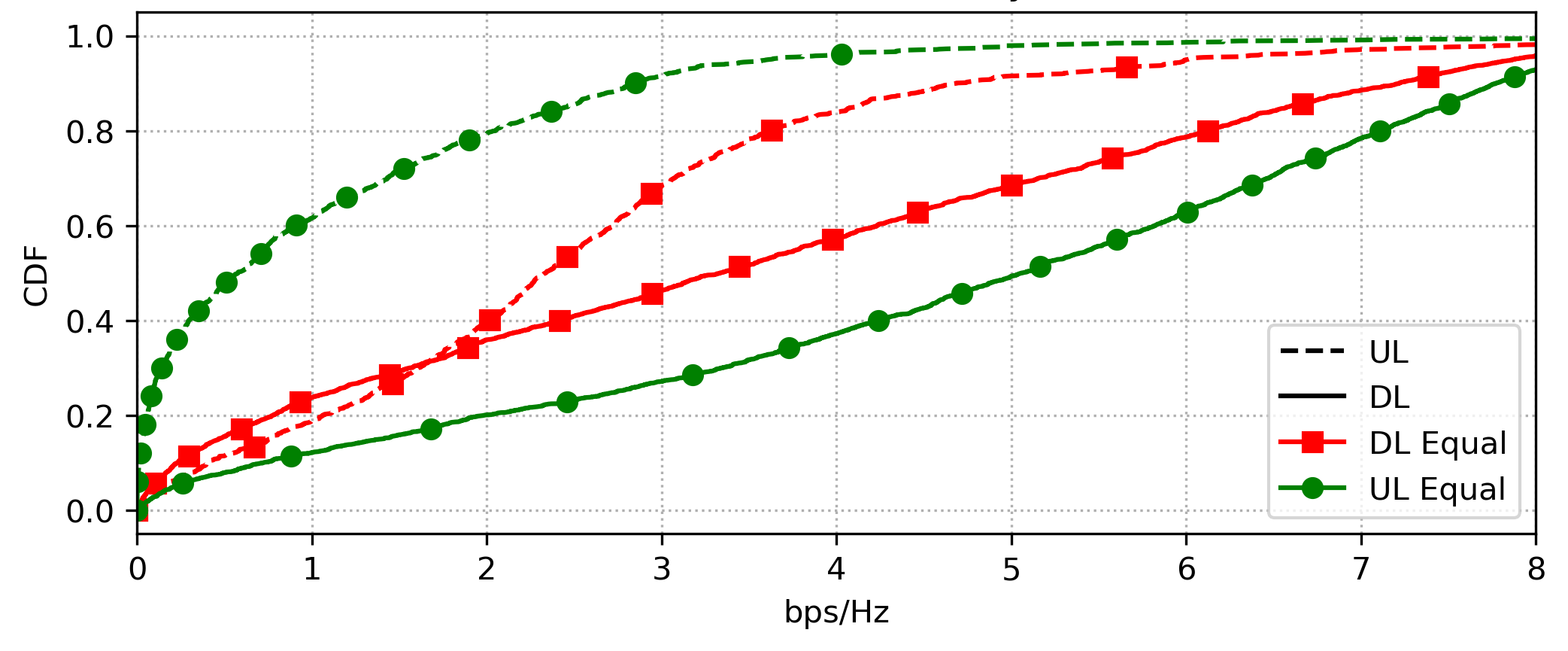}
	\caption{\ac{SE} \ac{CDF} of the \ac{PN} using the DL Heavy pattern when the \ac{SN} uses the DL or UL Equal pattern.}
	\label{fig:sepnedleupcomp}
	\vspace{-0.4cm}
\end{figure}

\FigRef{fig:sepndlhcomp} shows the performance of the \ac{PN} using the DL Heavy pattern, while the \ac{SN} interferes with each pattern of  \FigRef{fig:pat}. In terms of \ac{DL}/\ac{UL} proportion, as the number of interfering \ac{DL} symbols increases, the worst is the \ac{PN} performance. Moreover, looking at the positions of \ac{DL} and \ac{UL} slots on the interfering network, we observe that when the \ac{SN} uses the patterns with \ac{DL} symbols preceding \ac{UL}, there is an improvement of the lower \ac{SE} values, while when the \ac{UL} symbols precede the \ac{DL}, the higher \ac{SE} values are enhanced. This behavior is depicted in \FigRef{fig:comp}, where the values of the 15th and the 85th percentiles of the curves in \FigRef{fig:sepndlhcomp} are isolated.  \FigRef{fig:sepnedleupcomp} shows this effect in more detail, with the separate performances of the \ac{UL} and the \ac{DL} of the \ac{PN} when the \ac{SN} is employing the \ac{UL} or the \ac{DL} Equal pattern. Even with the number of \ac{DL} and \ac{UL} symbols being the same, their performances are significanty different, proving that the trade-off displayed in \FigRef{fig:comp} has to do with how the \ac{SN} is interfering on the \ac{PN}. %
From \FigRef{fig:pat}, we see that when the \ac{SN} uses the \ac{UL} Equal pattern, most of its \ac{DL} interferes with the \ac{PN}'s \ac{UL}, resulting in a stronger perceived interference due to the higher power at the \ac{SN}'s \ac{DL}; the \ac{SN}'s \ac{UL} interferes with the \ac{PN}'s \ac{DL} with lower power, resulting in lighter interference and better performance. The opposite happens when the \ac{DL} Equal pattern is used by the \ac{SN}, which enhances the \ac{PN}'s \ac{UL} \ac{SE} and diminishes the \ac{SE} of the \ac{DL}. %

From these results, we can see that there is a performance trade-off where the \ac{PN} can use the strategy in \SecRef{strategy} to either prioritize the worst or the best \acp{UE}' \ac{SE}. Based on this choice and in the pattern it is currently using, the \ac{PN} indicates to the \ac{SN} whether the initial symbols of the pattern are of \ac{DL} or \ac{UL}. %

When the \ac{PN} is employing the \ac{DL} heavy pattern, we have that the possible outcomes with the patterns available for the \ac{SN} in \FigRef{fig:pat} are exposed in \FigRef{fig:sepndlhcomp}. Choosing to prioritize the performance of the worst \acp{UE}, the \ac{PN} indicates to the \ac{SN} that the first symbols must be of \ac{DL}; that way, the \ac{SN}, according to the demands of its \acp{UE}, selects one of the patterns attending the \ac{PN} request. 

\begin{figure}[h!]
	\centering
	\includegraphics[width=\linewidth]{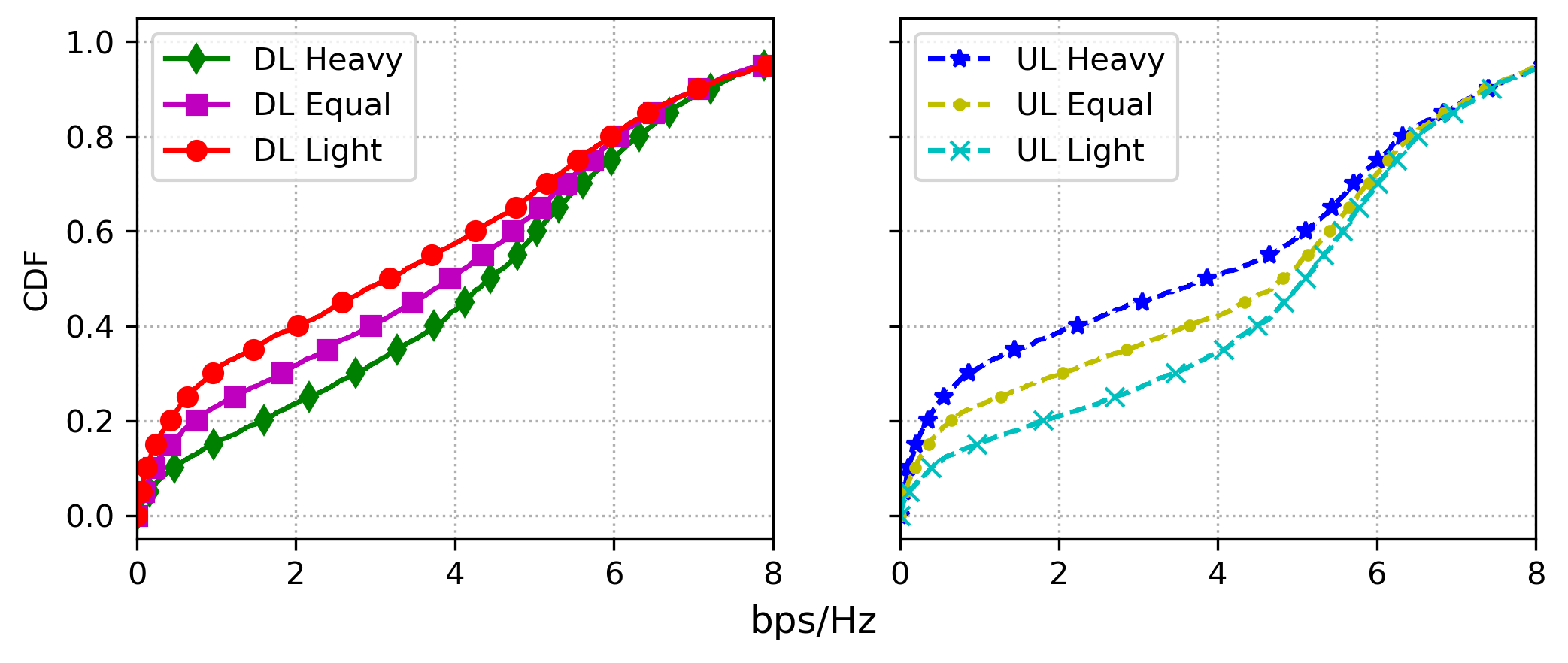}
	\caption{\ac{SE} \ac{CDF} of the \ac{SN} using each pattern in \FigRef{fig:pat} when the \ac{PN} uses the DL Heavy pattern.}
	\label{fig:sesncomp}
	\vspace{-0.3cm}
\end{figure}

\FigRef{fig:sesncomp} shows the performance of the \ac{SN} in this scenario. The figure presents the \ac{SE} \ac{CDF} of all \ac{UL} and \ac{DL} \acp{UE} for each of the patterns in \FigRef{fig:pat}. While there is no significant variance between the \ac{SE} curves, specially for the worst \acp{UE}, the performance of the best \acp{UE} is better when the employed pattern starts with the \ac{UL} symbols, hence, if the \ac{PN} demands that the \ac{SN} uses the patterns that start with the \ac{DL} symbols, there will be already a loss of performance. Considering only the allowed patterns, the \ac{SE} results improve as the number of \ac{DL} symbols increases. Considering their traffic demands and interference restrictions that the \ac{PN} might add in its request, however, the performance of the system might be severely impaired. One way to soothe this impact is to make more patterns available for the networks, as well as make use of the flexible symbols to increase the quality for its \acp{UE}.

\section{Conclusions}\label{Conclusions}

This work explored the coexistence of a \ac{D-MIMO} and a massive \ac{MIMO} network under \ac{TDD} and \ac{UL}/\ac{DL} asynchronization. It was seen that the position of the symbols and their proportion in the pattern have a significant influence on the networks performances, as the type of interference changes. With the slot coordination strategy, the \ac{PN} can manage its performance by sending requests for the \ac{SN} to choose a more convenient pattern. Future works might include an expansion of the presented strategy, with more slot patterns, the use of flexible symbols and with variation of the traffic load. 
\balance

\printbibliography

\end{document}